\documentclass[useAMS,usenatbib]{mn2e}
\usepackage{graphicx}

\title[Acoustic instability in the neutral precursor]{Acoustic instability in the neutral precursor region of collisionless shocks propagating into partially ionized plasmas}
\author[Y. Ohira]{Yutaka Ohira\thanks{E-mail:ohira@phys.aoyama.ac.jp}\\
Department of Physics and Mathematics, Aoyama Gakuin University, 
5-10-1 Fuchinobe, Sagamihara 252-5258, Japan
}
\begin{document}

\date{Accepted 2014 February 4.}

\pagerange{\pageref{firstpage}--\pageref{lastpage}} \pubyear{2014}

\maketitle

\label{firstpage}

\begin{abstract}
Recent studies about collisionless shocks in partially ionized plasmas 
showed that some of neutral particles leak into the shock upstream region 
from the downstream region. 
In this paper, we perform a linear analysis and show that acoustic waves 
are unstable in the neutral precursor region. 
The acoustic instability amplifies fluctuations of magnetic field and density in the upstream region. 
The fluctuations are indispensable for the diffusive shock acceleration and could be important for 
the downstream turbulence. 
\end{abstract}


\section{Introduction}

Galactic cosmic rays (CRs) are though to be accelerated by collisionless shocks 
driven by a supernova explosion \citep{axford77,krymsky77,bell78,blandford78}. 
Recent observations of supernova remnants (SNRs) support the CR acceleration 
in SNRs \citep{koyama95,ohiraetal11}. 
Although the second order acceleration by turbulence is a plausible acceleration 
mechanism for low energy CRs \citep{ohira13a}, 
the diffusive shock acceleration (DSA) is though to be the most plausible 
acceleration mechanism. 
In order to quickly accelerate particles, DSA needs strong magnetic field 
fluctuations compared to that in interstellar medium. 
Interestingly, it has been shown that particles accelerated by DSA can amply 
the magnetic field fluctuations in the shock upstream region \citep[for a recent review, see][]{schure12}. 
Although there are many studies about the magnetic field amplification in the shock 
upstream region, saturation of the amplification is still an open problem. 

The interstellar medium around SNRs is not always completely ionized. 
Ions produced by ionization of neutral particles excite several plasma instabilities 
and amplify the magnetic field fluctuations \citep{raymond08,ohira09,ohira10}. 
In addition, \citet{blasi12,ohira12} proposed that some of downstream 
neutral particles leak into the upstream region and the leaking neutral particles 
change the structure of shocks propagating into partially ionized plasmas. 
\citet{ohira13b} showed the leakage of neutral particles 
into the upstream region by hybrid particles simulations. 
Moreover, \citet{ohira13b} observed plasma instabilities in the upstream 
and downstream regions. 
Interestingly, the observed mode in the upstream region (neutral precursor region) 
is not any modes expected from kinetic instabilities, but the fast magnetosonic mode. 
The physical mechanism of the instability that excites the fast magnetosonic mode was not 
discussed in \citet{ohira13b}. 
Therefore, we investigate the excitation mechanism of the fast magnetosonic 
mode observed in the hybrid simulation of \citet{ohira13b} in this paper. 

We first summarize the simulation results of \citet{ohira13b} and provide a background plasma condition for a linear analysis (section \ref{sec:2}). Then, we perform the linear analysis of the fast magnetosonic mode with ionization of the leaking neutral particles (section \ref{sec:3}) and discuss the physical mechanism of the instability and impacts on the magnetic field amplification (section \ref{sec:4}). 

\section[]{Background condition in the neutral precursor region}
\label{sec:2}

In this section, we summarize simulation results of \citet{ohira13b}, 
providing a background condition for a linear analysis 
in the neutral precursor region. 
\citet{ohira13b} performed a two-dimensional hybrid simulation in order to investigate 
a perpendicular collisionless shock propagating into partially ionized plasmas. 
The simulation solved motion of protons and hydrogen atoms as particles, 
Maxwell's equations, collisional ionization and charge exchange. 
It was shown by the simulation that some of downstream neutral particles leak into 
the shock upstream region, the leaking neutral particles are ionized and 
become pickup ions in the upstream region, the pickup ions are preferentially accelerated, 
and the upstream flow is decelerated by the pickup ions. 
In the upstream rest frame, the mean velocity of leaking neutral particles is of the order of the shock velocity. 
In addition, the fast magnetosonic mode is excited in the neutral precursor region. 
The wavelength of the observed mode is about the gyro radius of the pickup ions, 
that is much smaller than the precursor length scale. 
The propagation direction and the amplitude are the direction of the shock normal 
and $\delta B/B_0 \approx \delta \rho /\rho_0 \approx 0.5$, respectively. 

After the leaking neutral particles are ionized, 
the ionized particles start to gyrate around the upstream magnetic fields with a 
relative velocity between the upstream flow and the leaking neutral particles. 
Then, ionized particles become hot pickup ions. 
The pickup ions have the pressure anisotropy, that is, 
the pressure perpendicular to the magnetic field line is larger than 
that parallel to the magnetic field line. 
The pressure anisotropy can excite the slow magnetosonic mode 
and the Alfv{\'e}n mode \citep{raymond08}, 
but cannot excite the fast magnetosonic mode observed in the simulation. 
The Drury instability can excite the fast magnetosonic mode if there are diffusive 
particles and the pressure gradient of the diffusive particles \citep{drury86,chalov88}.  
However, the simulation results showed that there is no diffusive particle 
in the upstream region. 

In this paper, we consider magnetohydrodynamic equations 
with injection of a hot plasma due to collisional ionization of the leaking neutral particles. 
We consider only collisional ionization because essential effects of charge exchange are the same as that of collisional ionization. 
We use Cartesian coordinates in which the shock normal is along the $x$ axis. 
We consider a one-dimensional flow in the $x$ direction 
and the magnetic field in the $z$ direction, 
where all values depend on position, $x$, and time, $t$.
Then, basic equations are given by
\begin{equation}
\frac{\partial}{\partial t} \rho + \frac{\partial}{\partial x} \left(\rho u\right) = \rho_{\rm n} \nu_{\rm i} ~~,
\label{eq:mass}
\end{equation}
\begin{equation}
\frac{\partial}{\partial t} \left(\rho u\right) + \frac{\partial}{\partial x} \left(\rho u^2+P +\frac{B^2}{8\pi}\right)= \rho_{\rm n} u_{\rm n} \nu_{\rm i} ~~,
\label{eq:momentum}
\end{equation}
\begin{eqnarray}
&&\frac{\partial}{\partial t} \left(\frac{1}{2}\rho u^2+\frac{P}{\gamma-1}+\frac{B^2}{8\pi}\right) \nonumber \\
&&+ \frac{\partial}{\partial x} \left(\frac{1}{2}\rho u^3+\frac{\gamma}{\gamma-1}Pu +\frac{B^2}{4\pi}u\right)= \frac{1}{2} \rho_{\rm n} u_{\rm n}^2 \nu_{\rm i} ~~,
\label{eq:energy}
\end{eqnarray}
\begin{equation}
\frac{\partial}{\partial t} B + \frac{\partial}{\partial x} \left(Bu\right) = 0 ~~,
\label{eq:induction}
\end{equation}
\begin{equation}
\frac{\partial}{\partial t} \rho_{\rm n} + \frac{\partial}{\partial x} \left ( \rho_{\rm n}u_{\rm n} \right ) = -\rho_{\rm n}\nu_{\rm i} ~~,
\label{eq:neutral}
\end{equation}
where $\rho, \rho_{\rm n}, u, u_{\rm n}, P, \gamma$ and $B$ are the density of plasma, the density of leaking neutral particles, the velocity of plasma, the velocity of leaking neutral particles, the pressure, the adiabatic index of plasma, and the magnetic field, respectively. 
The velocity of the leaking neutral particles, $u_{\rm n}$, can be assumed to be constant  because they do not interact with the electromagnetic field and the momentum change is negligible. 
The ionization frequency, $\nu_{\rm i}$, is given by 
\begin{equation}
\nu_i = \frac{\rho}{m} u_{\rm rel}\sigma_{\rm i}(u_{\rm rel}) ~~,
\end{equation}
where $\rho/m, u_{\rm rel}=|u_{\rm n} - u|$, and $\sigma_{\rm i}(u_{\rm rel})$ are the number density of ions, the relative velocity between plasma and the leaking neutral particles, and the cross section of collisional ionization, respectively. 
For simplicity, in equation (\ref{eq:energy}), we assume that leaking neutral particles are cold, 
but this assumption is not crucial. 

Strictly speaking, the mean profiles of plasma density and velocity slightly change 
in the neutral precursor region but they are almost constant 
because the leaking flux is only a few percent of the flux of the upstream flow in the shock rest frame \citep{ohira13b}. 
However, the pressure significantly changes in the neutral precursor region for high Mach number shock \citep{ohira12}. 
In addition, we consider perturbations with a length scale smaller than that of the neutral precursor, 
$\sim u_{\rm n}\nu_{\rm i}^{-1}$. 
Therefore, for simplicity, we assume that background fields are spatially uniform 
and constant with time except for the plasma pressure, $P$. 
Then, the background uniform solutions in the upstream rest frame are as follows.

\begin{eqnarray}
\rho &=& \rho_0~~,~\rho_{\rm n}=\rho_{\rm n,0}~,\\
u &=& 0~~,~u_{\rm n}=u_{\rm n,0}~,\\
B &=& B_0~~,\\
\nu_{\rm i,0}&=&\frac{\rho_0}{m}u_{\rm n,0}\sigma_{\rm i}(u_{\rm n,0})~~,\\
P(t) &=& P_0+\frac{\gamma-1}{2}\rho_{\rm n,0} u_{\rm n,0}^2\nu_{\rm i,0}t ~~.
\end{eqnarray}
The phase velocity of the fast magnetosonic mode is given by
\begin{eqnarray}
u_{\rm ph}(t) &=& \sqrt{\gamma \frac{P(t)}{\rho_0}+ \frac{B_0^2}{4\pi \rho_0} } \nonumber \\
&=&u_{\rm ph,0}\sqrt{1+\frac{t}{t_{\rm c}}}~~,
\end{eqnarray}
where $u_{\rm ph,0}=u_{\rm ph}(0)$ and $t_{\rm c}$ is given by
\begin{equation}
t_{\rm c} = \frac{2}{\gamma \left(\gamma-1\right)\xi}\left(\frac{u_{\rm ph,0}}{u_{\rm n,0}} \right)^2\nu_{\rm i,0}^{-1}~~,
\label{eq:tc}
\end{equation}
where $\xi=\rho_{\rm n,0}/\rho_0$ is the density ratio and less than about $0.1$, that depends on the shock velocity and the electron temperature in the shock downstream region. 
The pressure and the phase velocity can be assumed to be constant for $t < t_{\rm c}$. 
Note that $t_{\rm c}$ is much smaller than the ionization time scale, $\nu_{\rm i,0}^{-1}$, for high Mach number shocks 
because the velocity of leaking neutral particles, $u_{\rm n,0}$, is of the order of the shock velocity in the upstream rest frame. 

\section[]{Linear analysis}
\label{sec:3}

In this section, we perform a linear analysis of equations (\ref{eq:mass})-(\ref{eq:induction}) by using the background solutions. 
We do not consider perturbations of the leaking neutral particles. 
They are advected with the velocity of $u_{\rm n,0}$. 
On the other hand, perturbations of plasma propagate with the phase 
velocity smaller than $u_{\rm n,0}$, so that we can approximate that 
the perturbations of plasma are decoupled from perturbations of 
the leaking neutral particles. 
Then, the linearized perturbation equations are given by 
\begin{equation}
\frac{\partial}{\partial t} \delta\rho + \rho_0 \frac{\partial}{\partial x} \delta u = \rho_{\rm n,0} \left( \frac{d\nu_{\rm i,0}}{d\rho_0}\delta \rho - \frac{d\nu_{\rm i,0}}{du_{\rm n,0}}\delta u \right)~~,
\label{eq:per_mass}
\end{equation}
\begin{equation}
\rho_0 \frac{\partial}{\partial t} \delta u + \frac{\partial}{\partial x} \delta P +\frac{B_0}{4\pi}\frac{\partial}{\partial x} \delta B= \rho_{\rm n,0} u_{\rm n,0}\left( \frac{d\nu_{\rm i,0}}{d\rho_0}\delta \rho - \frac{d\nu_{\rm i,0}}{du_{\rm n,0}}\delta u \right) ~~,
\end{equation}
\begin{eqnarray}
\frac{1}{\gamma-1}\frac{\partial}{\partial t} \delta P  &+& \frac{B_0}{4\pi}\frac{\partial}{\partial t}\delta B + \left( \frac{\gamma}{\gamma-1}P(t) +\frac{B_0^2}{4\pi}\right) \frac{\partial}{\partial x} \delta u \nonumber \\
&=& \frac{1}{2}\rho_{\rm n,0}u_{\rm n,0}^2 \left( \frac{d\nu_{\rm i,0}}{d\rho_0}\delta \rho - \frac{d\nu_{\rm i,0}}{du_{\rm n,0}}\delta u \right)~~,
\label{eq:per_energy}
\end{eqnarray}
\begin{equation}
\frac{\partial}{\partial t} \delta B + B_0\frac{\partial}{\partial x} \delta u = 0 ~~.
\label{eq:per_induction}
\end{equation}

The above equations (\ref{eq:per_mass})-(\ref{eq:per_induction}) can be reduced to 
\begin{eqnarray}
0=\left\{ \frac{d^3}{d t^3} \right.&+&u_{\rm ph}(t)^2 k^2 \frac{d}{d t} \nonumber \\
&-& \xi \nu_{\rm i,0}\left(1-\frac{d \log \nu_{\rm i,0}}{d\log u_{\rm n,0}}\right)\frac{d^2}{d t^2} \nonumber \\
&+& i\xi \nu_{\rm i,0}u_{\rm n,0}k\left(1-\frac{\gamma-1}{2}\frac{d \log \nu_{\rm i,0}}{d\log u_{\rm n,0}}\right)\frac{d}{dt}  \nonumber \\
&-& \left. \xi \nu_{\rm i,0} \left(u_{\rm ph}(t)^2 - \frac{\gamma^2-1}{2}u_{\rm n,0}^2 \right)k^2  \right \}\delta u  ~~,
\label{eq:ode}
\end{eqnarray}
where the perturbation is assumed to be $\delta u\propto \exp(ikx)$.
Note that $ -1\la d \log \nu_{\rm i,0}/d\log u_{\rm n,0} \la 1$ for the shock velocity of the order of $2000~{\rm km/s}$ \citep{heng07}. 
Furthermore, the above equation does not explicitly contain the magnetic field and it contributes only the phase velocity. 
Therefore, the magnetic field is not directly important to the evolution of the perturbation. 

\subsection{For $t < t_{\rm c}$}
\label{sec:3.1}
\begin{figure}
\begin{center}
\includegraphics[width=80mm]{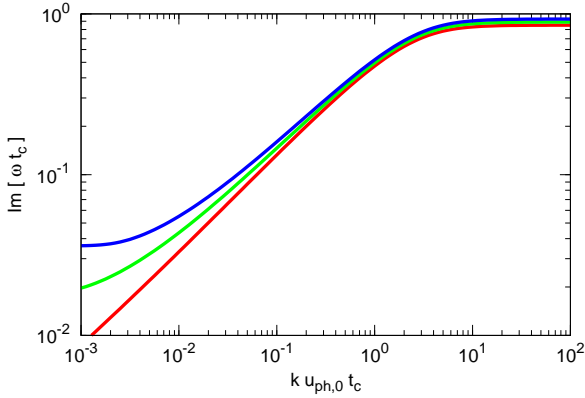}
\end{center}
\caption{Growth rate of the acoustic instability for $u_{\rm n,0}/u_{\rm ph,0}=10$ and $\gamma=5/3$. 
The red, green, and blue lines show the growth rates of the most unstable mode 
for $d\log \nu_{\rm i,0}/d\log u_{\rm n,0}=1,~0$, and $-1$, respectively.}
\label{fig:1}
\end{figure}
\begin{figure}
\begin{center}
\includegraphics[width=80mm]{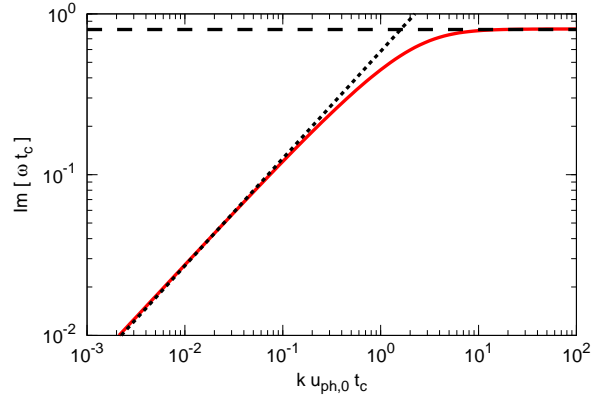}
\end{center}
\caption{The same as Fig.~\ref{fig:1} but for $u_{\rm n,0}/u_{\rm ph,0}=100$. 
The red solid line shows the growth rate of the most unstable mode for $d\log \nu_{\rm i,0}/d\log u_{\rm n,0}=1$. 
The black dotted and dashed lines show the asymptotic solution of the growth rate
for $k<k_{\rm c}$ and $k>k_{\rm c}$, respectively.}
\label{fig:2}
\end{figure}

For $t < t_{\rm c}$, the phase velocity, $u_{\rm ph}(t)$, can be assumed to be 
constant, $u_{\rm ph,0}$, so that the perturbation can be assumed to be $\delta u\propto \exp[i(kx-\omega t)]$. 
Then, the dispersion relation of equation (\ref{eq:ode}) is given by
\begin{eqnarray}
0=\omega^3 &-&u_{\rm ph,0}^2 k^2 \omega \nonumber  \\
&-&i\xi \nu_{\rm i,0}\left(1-\frac{d \log \nu_{\rm i,0}}{d\log u_{\rm n,0}}\right)  \omega^2  \nonumber \\
&-&i\xi \nu_{\rm i,0}u_{\rm n,0}k\left(1-\frac{\gamma-1}{2}\frac{d \log \nu_{\rm i,0}}{d\log u_{\rm n,0}}\right)\omega \nonumber \\
&+&i \xi \nu_{\rm i,0} \left(u_{\rm ph,0}^2 - \frac{\gamma^2-1}{2}u_{\rm n,0}^2 \right)k^2  ~~.
\label{eq:dispersion}
\end{eqnarray}
Three solutions of the dispersion relation 
($\omega=\omega(k)$) can be analytically obtained by using the Cardano's method. 
Then, we find that the fast magnetosonic modes propagating parallel and antiparallel to the leaking neutral particles 
are unstable, and the mode propagating parallel to the leaking neutral particle is the most unstable mode.

Fig.~\ref{fig:1} shows the growth rates of the most unstable mode for $u_{\rm n,0}/u_{\rm ph,0}=10$ and $\gamma = 5/3$. 
The red, green, and blue lines show the growth rates for $d\log \nu_{\rm i,0}/d\log u_{\rm n,0}=1,~0$, and $-1$, respectively. 
The growth rates increase with the increasing the wavenumber. 
It should be noted that the dispersion relation is not valid for a length scale smaller than  
the gyroradius of pickup ions and for a time scale smaller than the gyroperiod of pickup ions 
because the fluid approximation is not valid for the pickup ions. 

Fig.~\ref{fig:2} is the same as Fig.~\ref{fig:1} but for $u_{\rm n,0}/u_{\rm ph,0}=100$, that is, for high Mach number shocks. 
We show the growth rate only for $d\log \nu_{\rm i,0}/d\log u_{\rm n,0}=1$ because 
the growth rate does not significantly depend on $d\log \nu_{\rm i,0}/d\log u_{\rm n,0}$ for $u_{\rm n,0}/u_{\rm ph,0}\gg1$.  
For high Mach number shock ($u_{\rm n,0}\gg u_{\rm ph,0}$), 
the first, second, and final terms are dominant in equation (\ref{eq:dispersion}), 
so that the final term is the driving term of the acoustic instability. 
The driving term originates from the first term of the right hand side of equation~(\ref{eq:per_energy}). 
Therefore, the driving force is the energy injection due to ionization of leaking neutral particles. 
By considering only the three terms, 
the asymptotic analytical solution of the growth rate is given by 
\begin{eqnarray}
\rm{Im}[\omega]=\frac{\gamma+1}{2\gamma t_{\rm c}} \times \left \{ \begin{array}{ll}
\left( \frac{k}{k_{\rm c}}\right)^{2/3} & ~~({\rm for}~k <k_{\rm c}) \\
1 & ~~({\rm for}~k >k_{\rm c}) \\
\end{array} \right.~~,
\end{eqnarray}
where the characteristic wavenumber $k_{\rm c}$ is given by
\begin{eqnarray}
k_{\rm c} &=& \frac{\gamma+1}{\gamma t_{\rm c}u_{\rm ph,0}} \nonumber \\
&=& \frac{\xi \left(\gamma^2-1\right)}{2}\left(\frac{u_{\rm n,0}}{u_{\rm ph,0}}\right)^3 \frac{\nu_{\rm i}}{\Omega_{c}}r_{\rm g}^{-1}~~,
\end{eqnarray}
where $\Omega_{\rm c}$ and $r_{\rm g}=u_{\rm n,0}/\Omega_{\rm c}$ are the cyclotron frequency 
and the gyroradius of pickup ions, respectively. 
$\nu_{\rm i}/\Omega_{\rm c}\approx 10^{-5}$ and $u_{\rm n,0}/u_{\rm ph,0}\approx10^2 $ 
for young SNRs in the typical interstellar medium, and $\xi \la 0.1$, 
so that the length scale of $k_{\rm c}$ is comparable to or larger than the gyroradius of pickup ions, 
$r_{\rm g}$, for young SNRs. 

The shortest growth time scale is of the order of $t_{\rm c}$. 
However, equation (\ref{eq:dispersion}) is valid for $t < t_{\rm c}$, so that the exponential growth 
cannot be expected for a long time. 
We calculate a long time evolution in the next subsection. 

\subsection{For $t > t_{\rm c}$}
\label{sec:3.2} 

\begin{figure}
\begin{center}
\includegraphics[width=80mm]{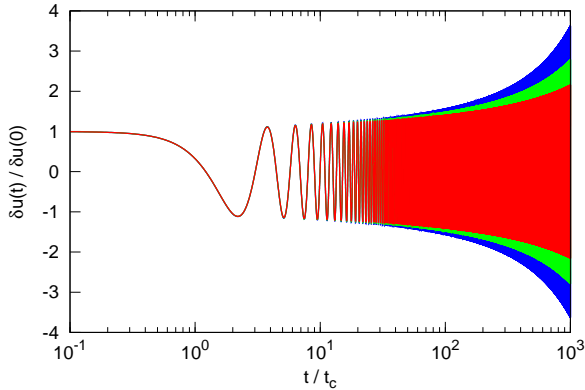}
\end{center}
\caption{Time evolution of the velocity perturbation at $x=0$ for $ku_{\rm ph,0}t_{\rm c}=1$, $u_{\rm n,0}/u_{\rm ph,0}=100$, and $\gamma = 5/3$. 
The red, green, and blue lines show the numerical results for $d\log \nu_{\rm i,0}/d\log u_{\rm n,0}=1,~ 0$, and $-1$, respectively.}
\label{fig:3}
\end{figure}
\begin{figure}
\begin{center}
\includegraphics[width=80mm]{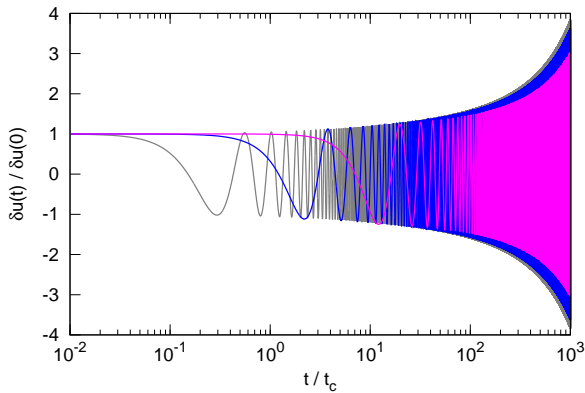}
\end{center}
\caption{Time evolution of the velocity perturbation at $x=0$ for $d\log \nu_{\rm i,0}/d\log u_{\rm n,0}=-1$, $u_{\rm n,0}/u_{\rm ph,0}=100$, and $\gamma = 5/3$. 
The grey, blue, and magenta lines show the numerical results for $ku_{\rm ph,0}t_{\rm c}=10,~1$, and $0.1$, respectively. }
\label{fig:4}
\end{figure}

In order to understand a long time evolution of perturbations, 
we solve equation (\ref{eq:ode}) by numerical simulations. 
Fig.~\ref{fig:3} shows the numerical results for $ku_{\rm ph,0}t_{\rm c}=1$, $u_{\rm n,0}/u_{\rm ph,0}=100$, and $\gamma=5/3$. 
The red, green, and blue lines show the results for $d\log \nu_{\rm i,0}/d\log u_{\rm n,0}=1,~0$, and $-1$, respectively. 
We adopt a normal fast magnetosonic mode propagating parallel to the leaking neutral particles 
as an initial perturbation, that is, $\delta u\propto \exp[i(kx-ku_{\rm ph,0}t)]$.
The transit time in the neutral precursor region is about the ionization time scale, $\nu_{\rm i}^{-1}$, 
that is about $10^3~t_{\rm c}$ for $u_{\rm n,0}/u_{\rm ph,0}=100$.  

The amplitudes slowly increase with time.
Therefore, the fast magnetosonic mode is still unstable even when $t>t_{\rm c}$. 
Three lines are almost the same in the early phase, but in the late phase, 
the amplitude for the case of $d\log \nu_{\rm i,0}/d\log u_{\rm n,0}=-1$ grows slightly faster than 
that for the other cases. 
These behaviors can be understood from the dispersion relations shown in Figs. (\ref{fig:1}) and (\ref{fig:2}). 
The dispersion relation does not depend on the velocity dependence of the ionization frequency, $d\log \nu_{\rm i,0}/d\log u_{\rm n,0}$ for a large $u_{\rm n,0}/u_{\rm ph}$, but it depends on $d\log \nu_{\rm i,0}/d\log u_{\rm n,0}$ for a small $u_{\rm n,0}/u_{\rm ph}$. 
The phase velocity, $u_{\rm ph}(t)$, increases with time for $t>t_{\rm c}$.
Therefore, the evolution of the amplitude depends on $d\log \nu_{\rm i,0}/d\log u_{\rm n,0}$ in the late phase.  

Fig. \ref{fig:4} shows the numerical results for $d\log \nu_{\rm i,0}/d\log u_{\rm n,0}=-1$, $u_{\rm n,0}/u_{\rm ph,0}=100$, and $\gamma=5/3$. 
The grey, blue, and magenta lines show the numerical results for $ku_{\rm ph,0}t_{\rm c}=10,~1$, and $0.1$, respectively. 
The evolution hardly depends on the wavenumber for $k>k_{\rm c}\approx (u_{\rm ph,0}t_{\rm c})^{-1}$, 
but the growth rates increase with the increasing the wavenumber for $k<k_{\rm c}$. 
These behaviors are also the same as that of the dispersion relations shown in Figs. (\ref{fig:1}) and (\ref{fig:2}). 
Therefore, for the young SNRs, the mode with $k= k_{\rm c} \approx r_{\rm g}^{-1}$ has the maximum growth rate even when $t>t_{\rm c}$. 
We again note that the numerical solutions are not valid for a wavelength smaller than the gyroradius of pickup ions because the fluid approximation is not valid for the pickup ions.

\section[]{Discussion}
\label{sec:4}

First, we discuss the physical mechanism of the acoustic instability. 
For $t<t_{\rm c}$, as mentioned in the section 3.1, 
the driving force is the energy injection due to ionization of leaking neutral particles.
The first term of the right hand side of equation (\ref{eq:per_energy}) means that 
leaking neutral particles are more ionized in denser regions 
and their kinetic energy is more injected in the denser regions but the mass injection due to ionization is negligible. 
Then, the pressure gradient between low density regions and high density regions becomes large compared with 
that before injection of ionized particles. 
Therefore, the amplitude of the acoustic wave grows with oscillation. 

For $t>t_{\rm c}$, the phase velocity, $u_{\rm ph}(t)$, becomes comparable to the velocity of 
leaking neutral particles, $u_{\rm n,0}$, so that all the terms in equation (\ref{eq:ode}) become comparable.
Therefore, in addition to the energy injection due to ionization, the mass and momentum injection due to ionization 
also contributes the acoustic instability. 
Furthermore, not only the density perturbation but also the velocity perturbation (the second terms of the right hand side of equations (\ref{eq:per_mass})-(\ref{eq:per_energy})) contributes the acoustic instability.

The acoustic instability could be important for other instabilities 
(e.g. the Drury instability and parametric instabilities) and the magnetic field amplification in the CR precursor \citep{beresnyak09,drury12} and downstream regions \citep{giacalone07,inoue09,guo12} 
as a seed of perturbations with a small length scale. 

We did not take into account gradient of background velocity and density in the neutral precursor region. 
\citet{ohira13b} showed that the gradient is very small for the shock velocity of about $3000~{\rm km/s}$. 
For slower shock velocity, it is expected that more neutral particles leak into the upstream region and 
the gradient in the neutral precursor region becomes larger \citep{blasi12,ohira12}. 
In this case, a wave-action analysis like \citet{drury86} is needed to properly analyze the acoustic instability.
Furthermore, we did not take into account kinetic effects of pickup ions in this paper. 
These issues will be addressed in future work.

\section[]{Summary}
\label{sec:5}

In this paper, we have showed that the fast magnetosonic mode is unstable in the neutral precursor region 
where neutral particles are leaking from the downstream region. 
The acoustic instability has already been observed in a hybrid simulation for collisionless shocks 
propagating into partially ionized plasmas \citep{ohira13b}. 
In this paper, we provided the physical mechanism and linear analysis. 
The instability could be important for other instabilities as a seed of perturbations with a small length scale. 
\section*{Acknowledgments}

We thank M. A. Lee for useful discussion. 
This work is supported in part by grant-in-aid from the Ministry of Education, 
Culture, Sports, Science, and Technology (MEXT) of Japan, No.~24$\cdot$8344.

\label{lastpage}

\begin{thebibliography}{99}
%
\bibitem[\protect\citeauthoryear{Axford et al.}{1977}]{axford77}
Axford, W. I., Leer, E., \& Skadron, G., 1977, Proc. 15th Int. Cosmic Ray Conf., Plovdiv, 11, 132
%
\bibitem[\protect\citeauthoryear{Bell}{1978}]{bell78}
Bell, A. R., 1978, MNRAS, 182, 147
%
\bibitem[\protect\citeauthoryear{Blandford \& Ostriker}{1978}]{blandford78}
Blandford, R. D., \& Ostriker, J. P., 1978, ApJ, 221, L29
%
\bibitem[\protect\citeauthoryear{Blasi et al.}{2012}]{blasi12}
Blasi, P., Morlino, G., Bandiera, R., Amato, E., Caprioli, D., 2012, ApJ, 755, 121
%
\bibitem[\protect\citeauthoryear{Beresnyak et al.}{2009}]{beresnyak09}
Beresnyak, A., Jones, T. W., \& Lazarian, A., 2009, ApJ, 707, 1541
%
\bibitem[\protect\citeauthoryear{Chalov}{1988}]{chalov88}
Chalov, S. V., 1988, Sov. Astron. Lett., 14, 114
%
\bibitem[\protect\citeauthoryear{Drury \& Falle}{1986}]{drury86}
Drury, L. O'C., \& Falle, S. A. E. G., 1986, MNRAS, 223, 353
%
\bibitem[\protect\citeauthoryear{Drury \& Downes}{2012}]{drury12}
Drury, L. O'C., \& Downes, T. P., 2012, MNRAS, 427, 2308
%
\bibitem[\protect\citeauthoryear{Giacalone \& Jokipii}{2007}]{giacalone07}
Giacalone, J., \& Jokipii, J.R., 2007, ApJ, 663, L41
%
\bibitem[\protect\citeauthoryear{Guo et al.}{2012}]{guo12}
Guo, F., Li, S., Li, H., Giacalone, J., Jokipii, J. R., \& Li, D., 2012, ApJ, 747, 98
%
\bibitem[\protect\citeauthoryear{Heng \& McCray}{2007}]{heng07}
Heng, K. \& McCray, R., 2008, ApJ, 654, 923
%
\bibitem[\protect\citeauthoryear{Inoue et al.}{2009}]{inoue09}
Inoue, T., Yamazaki, R. \& Inutsuka, S., 2009, ApJ, 695, 825
%
\bibitem[\protect\citeauthoryear{Koyama et al.}{1995}]{koyama95}
Koyama, K., Petre, R., Gotthelf, E. V., Hwang, U., Matsuura, M., Ozaki, M., \& Holt, S. S., 1995, Nature, 378, 225
%
\bibitem[\protect\citeauthoryear{Krymsky}{1977}]{krymsky77}
Krymsky, G. F., 1977, Dokl. Akad. Nauk SSSR, 234, 1306
%
\bibitem[\protect\citeauthoryear{Ohira et al.}{2009b}]{ohira09}
Ohira, Y., Terasawa, T., \& Takahara, F., 2009b, ApJ, 703, L59
%
\bibitem[\protect\citeauthoryear{Ohira \& Takahara}{2010}]{ohira10}
Ohira, Y., \& Takahara, F., 2010, ApJ, 721, L43
%
\bibitem[\protect\citeauthoryear{Ohira et al.}{2011}]{ohiraetal11}
Ohira, Y., Murase, K. \& Yamazaki, R., 2011, MNRAS, 410, 1577
%
\bibitem[\protect\citeauthoryear{Ohira}{2012}]{ohira12}
Ohira, Y., 2012, ApJ, 758, 97
%
\bibitem[\protect\citeauthoryear{Ohira}{2013a}]{ohira13a}
Ohira, Y., 2013a, ApJ, 767, L16
%
\bibitem[\protect\citeauthoryear{Ohira}{2013b}]{ohira13b}
Ohira, Y., 2013b, Phys. Rev. Lett., 111, 245002
%
\bibitem[\protect\citeauthoryear{Raymond et al.}{2008}]{raymond08}
Raymond, J. C., Isenberg. P. A., \& Laming, J. M., 2008, ApJ, 682, 408 
%
\bibitem[\protect\citeauthoryear{Schure et al.}{2012}]{schure12}
Schure, K. M., Bell. A. R., Drury, L. O'C., \& Bykov, A.M., 2012, SSR, 173, 491 
\end{thebibliography}
\end{document}